\documentstyle[epsf,conf_iap,10pt]{article}
\newcommand{\la}{\mathrel{\mathchoice
{\vcenter{\offinterlineskip\halign{\hfil
$\displaystyle##$\hfil\cr<\cr\sim\cr}}}
{\vcenter{\offinterlineskip\halign{\hfil$\textstyle##$\hfil\cr
<\cr\sim\cr}}}
{\vcenter{\offinterlineskip\halign{\hfil$\scriptstyle##$\hfil\cr
<\cr\sim\cr}}}
{\vcenter{\offinterlineskip\halign{\hfil$\scriptscriptstyle##$\hfil\cr
<\cr\sim\cr}}}}}
\newcommand{\ga}{\mathrel{\mathchoice
{\vcenter{\offinterlineskip\halign{\hfil
$\displaystyle##$\hfil\cr>\cr\sim\cr}}}
{\vcenter{\offinterlineskip\halign{\hfil$\textstyle##$\hfil\cr
>\cr\sim\cr}}}
{\vcenter{\offinterlineskip\halign{\hfil$\scriptstyle##$\hfil\cr
>\cr\sim\cr}}}
{\vcenter{\offinterlineskip\halign{\hfil$\scriptscriptstyle##$\hfil\cr
>\cr\sim\cr}}}}}
\newcommand{\Lya}{\mbox{Ly-$\alpha$\,}}
\newcommand{\Msol}{\mbox{M$_\odot$}}
\begin{document}
\heading{The relation between QSO absorption systems and high redshift galaxies} 
\par\medskip\noindent
\author{Matthias Steinmetz}
\address{Steward Observatory, University of Arizona, Tucson, AZ 85721, USA}

\begin{abstract}
The relation between high redshift galaxies and QSO absorption systems is
discussed in the context of hierarchical galaxy formation.  It is demonstrated
that imprints of the reionization history of the universe are detectable in the
$b$-parameter distribution of the \Lya forest at redshifts two to four, favoring
models in which hydrogen and helium are reionized simultaneously at or before
redshift five by a quasar-like spectrum of UV photons. Hydrodynamical
simulations including star formation, feedback due to supernovae and chemical
enrichment are also presented. Energy feedback in form of kinetic energy can give
rise to an efficient transport of metals out to distances of a few hundred
kpc. The observational signature of this metal transport mechanism compared to simple
homogeneous enrichment models is discussed. Finally, it is shown that present day
$L^*$ galaxies typically have several progenitors at $z\approx3$ spread over
a few hundred kpc. These progenitors are closely
associated with damped \Lya and Ly-limit systems, and
have velocity dispersions, luminosities and colors comparable
to U-dropout galaxies at $z\approx 3$.
\end{abstract}
\section{Introduction}

Over the past few years gasdynamical simulations have had an enormous impact on our
theoretical understanding of QSO absorption systems. These simulations can
explain the basic properties of QSO absorbers
covering many orders of magnitude in column density 
(see, e.g., the many simulation related contributions in this volume). 
While the lowest column density systems ($\log N_{\rm HI} \approx 12-14$) 
arises from gas in voids and sheets of the ``cosmic web'', systems of higher
column density are produced by filaments ($\log N \approx 14-17$) or even gas
which has cooled and collapsed in virialized halos ($\log N > 17$).  
So far, numerical simulations have been  applied primarily to systems with lower 
column densities ($\log N \la 17$), corresponding
to gas densities below $10^{-2}\,$cm$^{-3}$. At these low
densities the important physical processes are
relatively simple and well understood. Fluctuations are still only mildly
non-linear and the gas is essentially in photoionization  equilibrium with the 
UV background. Cooling times are long compared to dynamical time scales.

These simulations have so far neglected the effects of star
formation and related feedback processes. They thus suffer
from the overcooling problem, which results in disk
galaxies which are too massive/luminous and too concentrated compared with
present day spiral galaxies \cite{NS}. This problem may be overcome by
including an efficient heating mechanism as, e.g., energy feedback from supernovae. Metals
ejected from star-forming galaxies in the process of formation also provide
an attractive scenario to explain the pollution of the IGM with metals. However,
shocks caused by this outflowing gas may also dramatically affect the
thermodynamics of the intergalactic medium.

In this contribution, I will present some examples how non-equilibrium thermodynamics
and star formation may affect properties of QSO absorbtion systems. In a first
application (section 2) the influence of the reionization history of the
universe on the thermodynamics of the IGM is discussed. Section 3 shows to what
extent winds driven by energy feedback due to supernovae can pollute the
intergalactic medium with metals. Section 4 addresses the relation between high
column density absorbers and the population of local and high redshift galaxies.

\section{Disentangling the reionization history of the universe}

\begin{figure}[t]
{\vskip0.2cm
\hskip0.04 \hsize\epsfxsize=0.9\hsize\epsffile{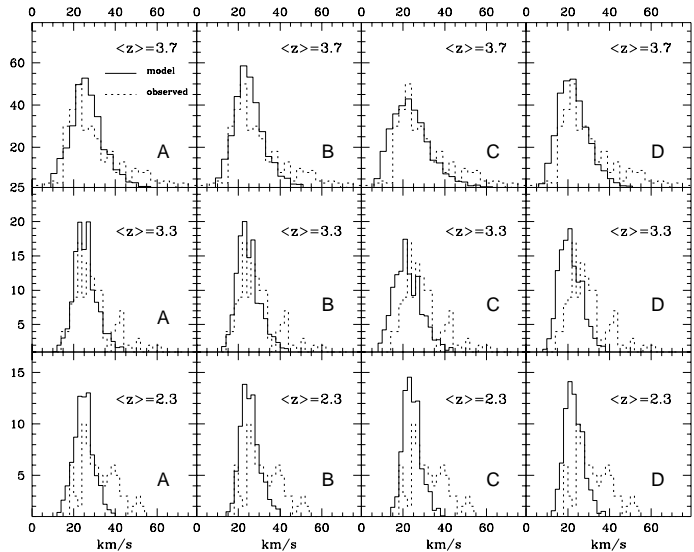}\hfill\break}
\vspace{-4.0cm}
\caption[]{Observed (dashed line) and simulated (solid line)
Doppler parameter distributions for models A--D (from left to right). 
The mean redshift of the samples is $\langle z\rangle$=2.3,3.3,3.7.}
\end{figure}

This section investigates how non-equilibrium thermodynamics may affect
properties of the \Lya forest, in particular the $b$-parameter distribution and
its dependence on the reionization history of the universe \cite{HS}.  During
reionization the IGM is heated up due to the residual energy of photons absorbed
by hydrogen and helium.  Cooling due to Compton scattering (at $z\ga 5$) and due
to adiabatic expansion continuously lowers the temperature
until photoheating and cooling are in equilibrium. During the epoch between
reionization and equilibrium, the temperature-density distribution of the IGM
depends on the exact balance between these heating and cooling processes.
Different reionization histories thus result in different $\varrho-T$
relations at low densities. These differences may be observable by comparing the
simulated and observed $N-b$ relation ($b$-parameter distribution). Such a
comparison requires to follow self-consistently the non-equilibrium evolution of
the baryonic species (H, H$^{+}$, He, He$^{+}$, He$^{++}$, and $e^-$). It is
further complicated by the fact the the $b$ parameter is typically dominated by
bulk motions rather than the IGM's temperature.

In order to demonstrate that different ionization histories can leave some
observable imprints in the $b$-parameter distribution, four different models 
have been simulated to cover  the uncertainty concerning the respective role of quasars 
and stars in the reionization of the Universe: 

\begin{list}{$\bullet$}{\setlength{\leftmargin}{16mm}
\setlength{\labelwidth}{17mm}}

\item[Model A:]
Ionizing background as proposed by Haardt \& Madau \cite{HM}. The spectrum is a power-law with 
$\alpha$=1.5  processed by the intervening \Lya absorption. 

\item[Model B:]
Same as model A but the redshift evolution is stretched towards higher redshift. 

\item[Model C:]
Same as model A but the redshift evolution is compressed towards lower redshift.
A stellar component is added to the UV background which reionizes hydrogen at
redshift $z\sim 6$ with a soft spectrum (power law with $\alpha$=5).

\item[Model D:]
Same as C but the stellar component  reionizes hydrogen at redshift 
$z\sim 30$.   

\end{list}

Model A was chosen by Haardt \& Madau \cite{HM} to represent the 
UV background due to observed quasars. Model B mimics the existence of
an as yet undetected population of quasar at redshifts beyond five.
Models C and D address the possibility that the Haardt \& Madau model
overestimates the UV background due to quasars at redshifts larger
than three. In the latter case the UV background would have to be 
dominated by a stellar  contribution at these redshifts. 

The simulations are performed using GRAPESPH \cite{S} a smoothed particle
hydrodynamics code which includes the relevant cooling and heating processes
and follows self-consistently the non-equilibrium evolution of the baryonic
species (H, H$^{+}$, He, He$^{+}$, He$^{++}$, and $e^-$). The Doppler parameters
were obtained by fitting Voigt profiles to artificial spectra with the automatic
line fitting program AUTOVP kindly provided by Romeel Dav\'e \cite{De}.

Figure 1 compares observed and simulated Doppler parameter distribution at three
different redshifts \cite{Lu,Co}.  The mean redshift and the column density
range of the sample of lines identified in the artificial spectra has been
matched to those of observed samples.  At high redshift ($\langle
z\rangle$=3.3,3.7) there is good agreement between the observed and simulated
Doppler parameter distribution in the case of models A and B. The only
discrepancy is a modest high-velocity tail which is present in the observed data
but not in the simulated distribution.  Models C and D clearly disagree with the
observed distribution.  They show a considerable fraction of lines which are
cooler than the lower cut-off of 15 km/s. At redshift 2.3 there is a significant
discrepancy between simulated and observed distribution for all four models.
The observed distribution shows a pronounced high-velocity tail which is not
reproduced by any of the models. The failure to reproduce the high-velocity tail
may be related to the too small a box size of the simulation which thus miss
larger-scale bulk motions \cite{HR}. An alternative interpretation is that
simulations significantly underestimate the fraction of the IGM in a hot
collisionally ionized phase because they do not include any feedback effects due
to star formation.

\section{Metal Enrichment of the IGM}

The simulations in this and the next section include star formation and the
effects of energy feedback due to supernovae. Star formation is modelled by a heuristic
scheme in which high density, rapidly cooling gas is transformed into stars
assuming a Miller-Scalo IMF. Stars more massive than 8\,\Msol\, are assumed to
explode as Type II supernovae and deposit locally $\approx 10^{51}$\,erg into
the ISM. A fraction $f_v$ (typically about 10\%)
of this energy is invested in modifying the
kinetic energy of the surrounding gas while the remaining $(1-f_v)$ heats up the
neighboring gas. Furthermore, supernovae  also enrich neighboring fluid elements
with metals. More details on the star formation algorithm are given in \cite{NW}
and \cite{SM}.

\begin{figure}[t]
{\hskip0.04 \hsize\epsfxsize=0.9\hsize\epsffile{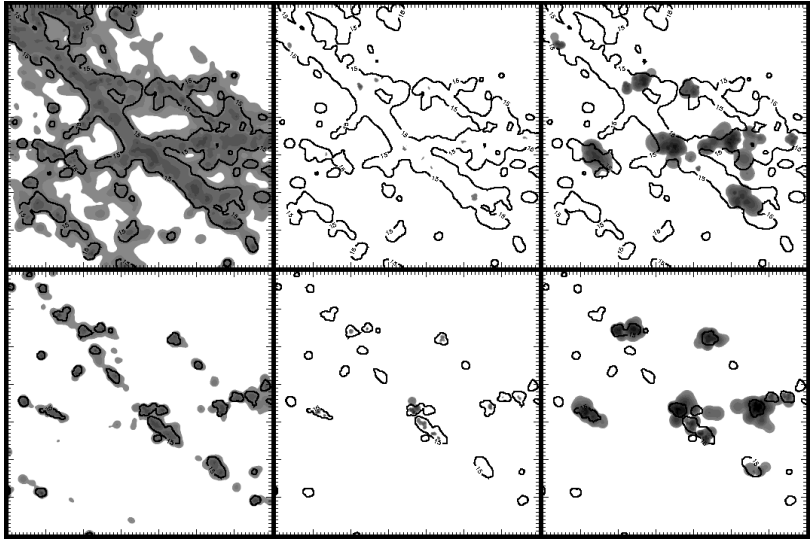}\hfill\break}
\caption[]{CIV column density distribution for $z=4$ (top row) and
$z=3$ (bottom row). The left column corresponds to a model assuming a constant,
homogeneously distributed metallicity ($[Z]=-2.5$), the middle column to
a model with feedback due to thermal energy ($f_v=0$), the right column to a model with
feedback due to thermal and turbulent energy ($f_v=0.1$). The solid contour line
encloses regions with a HI column density of above $\log N=15$, the gray shaded
region marks a CIV column density of $\log N > 11.7$. The size of each frame is
2.8\,Mpc comoving.}
\end{figure}

Figure 2 shows CIV column density maps for redshift 3 and 4 for three
different enrichment scenarios. The model shown in the
left panels assumes a homogeneous enrichment with a metallicity of $[Z] = -2.5$.
The properties of metal-line systems arising in such a model have been shown to
match observations fairly well \cite{RH,He}. It also predicts at $z=3$ a high
fraction of metal-line systems for HI column densities above $\log N=15$,
consistent with observations. In the simulations shown in the middle and right
panel, star formation and enrichment is treated self-consistently but with
different feedback parameters, $f_v= 0$ (pure thermal feedback) for the models
shown in the middle column and $f_v= 0.1$ for the models shown in the right
column. Models with $f_v=0.1$ have also been shown to reproduce the present day
galaxy luminosity function \cite{CF}.  In the thermal feedback model, the metal
distribution is very concentrated towards the center of galaxies where stars are
actually formed. Only a small fraction of LOS with $\log N(HI) > 15$ would show
an detectable CIV absorption, inconsistent with
observations. This flaw is cured in the turbulent feedback model, which
at $z\approx 3$ shows an absorption fraction similar to that of the homogeneous
model, though the correlation between HI and CIV absorption is weaker. Considering
the redshift evolution opens an interesting opportunity to discriminate between
these different enrichment scenarios. The area enclosed by the $\log N(HI) > 15$
contour increases substantially going from redshift 3 to 4. Similarly the area
with $\log N(CIV) > 11.7$ increases for the homogeneous enrichment
model. Contrary, this area is about constant, if not decreasing with redshift
for the two feedback models. Translated into the $dN/dz$ relation for CIV, the
homogeneous enrichment model predicts a strong increase with redshift while the
two feedback models predict $dN_{\rm CIV}/dz \approx const$.

It should be noted that these results are in partial disagreement with
simulation recently done by Gnedin \cite{Gn}, who finds 
a substantially weaker effect of supernovae feedback, but 
a stronger ejection of metals due to dynamics. While the differences in the
simulation including supernovae feedback can be likely accounted to the different
star formation and feedback models, the origin of the differences if supernovae
feedback is minimized is much less clear and requires further investigation.

\section{The connection between present day, galaxies, high-z galaxies and
QSO absorption systems}

\begin{figure}[t]
{\hskip0.0\hsize\epsfxsize=0.48\hsize\epsffile{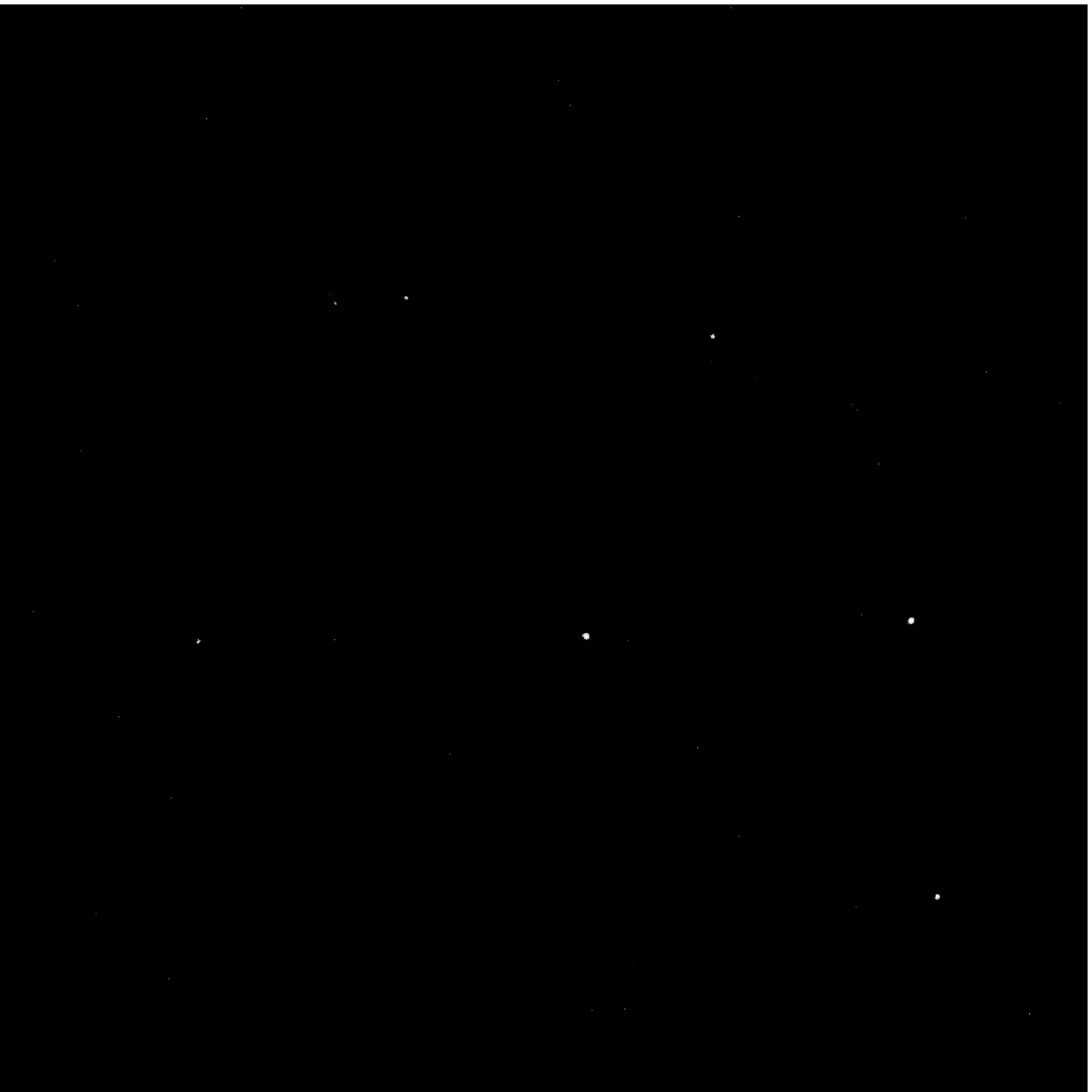}\hskip0.04\hsize\epsfxsize=0.48\hsize\epsffile{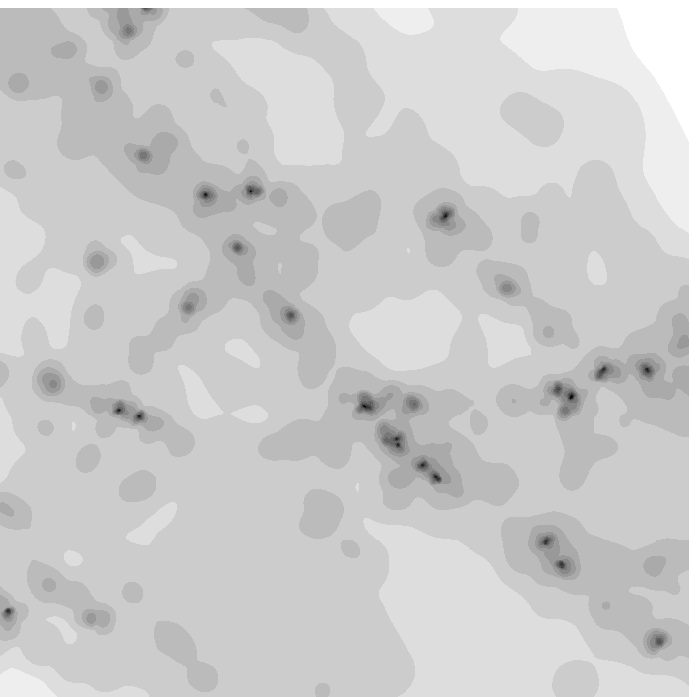}\break}
%
\caption[]{Left: Artificial I-band image of a galaxy in the process of formation
$z\approx 3$ in an HDF like exposure. Resolution, Noise, PSF and efficiency are
taken to match that of the HST WFPC2 camera.  Right: HI column density map at z=3.
Each frame has a sidelength of 2.8\,Mpc (comoving).}
\end{figure}

The simulation presented in the last section can also be used to 
investigate further the relation between QSO absorbers and galaxies in the process of
formation. Each ``star'' particle corresponds to
a population of about $10^6$ stars formed in a star-burst 
manner.  By means of spectro-photometric models \cite{Fr}, predictions for
the emission of these galaxies in the broad-band colors UBVRIK can be made.

In Figure 3, a HI column density map is compared with such an synthetic I-band
image of the stellar component. The I band image includes
noise, PSF and exposure time similar to that of the Hubble Deep Field. The
artificial image shows about 8 detectable protogalactic clumps (PGCs). Each of
these PGCs is situated close to a region of very high column density ($\log N >
17$). However, there is still a substantial number of Ly-limit and damped
\Lya systems, which do not host a stellar PGC.

Due to the small physical size of of the high resolution region of the
simulation, all PGCs share virtually the same redshift ($\Delta v\approx
400$\,km/sec).  In total a set of about 20 galaxies has been simulated with
circular velocities ranging between 100 and 200 km/s (at $z=0$). Each of these
galaxies gives rise to a couple of detectable ($I<26$) progenitors at $z\approx
2-3$, a behavior which nicely accounts for the increasing evidence for redshift
clustering at redshifts above 2 \cite{Wi,St}.

It is also of interest to take a closer look at further properties of these
PGCs.  Luminosity as well as  circular velocity of the most massive
clump at $z\approx 3$ is very similar to that of its present day
counterpart. The mass of the most massive progenitor is, however, only about
10\% of that at the present epoch. Star formation rates are fairly moderate,
typically a few to several solar masses per year. These characteristics are at
least qualitatively in very good agreement with properties of the recently
detected population of galaxies at $z=3$ and give some support for the hypothesis
that the U-dropout galaxies are the progenitors of the present population of
massive galaxies \cite{St}.

\acknowledgements{This article includes work from collaborations with
G.~Contardo, M.~Haehnelt, J.~Navarro and M.~Rauch.} 


%
\vfill
\end{document}